\documentclass[aps,pra,twocolumn,showpacs,superscriptaddress]{revtex4-1}
\usepackage[colorlinks ,linkcolor=blue,anchorcolor=blue,citecolor=blue,urlcolor=blue]{hyperref}
\usepackage{graphicx}
\usepackage{epstopdf}
\def\be{\begin{equation}} \def\ee{\end{equation}}
\def\be{\begin{equation}} \def\ee{\end{equation}}
\def\bea{\begin{eqnarray}} \def\eea{\end{eqnarray}}

\def\nn{\nonumber}

\begin{document}

\title{Quantum anomalous Hall state in a fluorinated 1T-MoSe$_{2}$ monolayer}
\author{Zhen Zhang}
\affiliation{School of Physics and Technology, Nanjing Normal University, Nanjing 210023, China}
\affiliation{Center for Quantum Transport and Thermal Energy Science (CQTES), Nanjing Normal University, Nanjing 210023, China}
\author{Zhichao Zhou}
\email{zczhou@njnu.edu.cn}
\affiliation{School of Physics and Technology, Nanjing Normal University, Nanjing 210023, China}
\author{Xiaoyu Wang}
\affiliation{School of Physics and Technology, Nanjing Normal University, Nanjing 210023, China}
\affiliation{Center for Quantum Transport and Thermal Energy Science (CQTES), Nanjing Normal University, Nanjing 210023, China}
\author{Huiqian Wang}
\affiliation{School of Physics and Technology, Nanjing Normal University, Nanjing 210023, China}
\affiliation{Center for Quantum Transport and Thermal Energy Science (CQTES), Nanjing Normal University, Nanjing 210023, China}
\author{Xiuling Li}
\affiliation{School of Physics and Technology, Nanjing Normal University, Nanjing 210023, China}
\affiliation{Center for Quantum Transport and Thermal Energy Science (CQTES), Nanjing Normal University, Nanjing 210023, China}
\affiliation{National Laboratory of Solid State Microstructures, Nanjing University, Nanjing 210093, China}
\author{Xiao Li}
\email{lixiao@njnu.edu.cn}
\affiliation{School of Physics and Technology, Nanjing Normal University, Nanjing 210023, China}
\affiliation{Center for Quantum Transport and Thermal Energy Science (CQTES), Nanjing Normal University, Nanjing 210023, China}

\begin{abstract}
The quantum anomalous Hall state with a large band gap and a high Chern number
is significant for practical applications in spintronics. 
By performing first-principles calculations, we investigate electronic properties of the fully fluorinated 1T-MoSe$_{2}$ monolayer. 
Without considering the spin-orbit coupling, the band structure demonstrates single-spin semi-metallic properties 
and the trigonal warping around $K_{\pm}$ valleys. 
The introduction of the spin-orbit coupling
opens considerable band gaps of $117.2$ meV around the two valleys, 
leading to a nontrivial quantum anomalous Hall state with a Chern number of $|C|=2$, 
which provides two chiral dissipationless transport channels from topological edge states 
and associated quantized anomalous Hall conductivity. 
In addition, an effective model is constructed to describe the low-energy physics of the monolayer.
Our findings in  the MoSe$_{2}$F$_{2}$ monolayer sheds light on large-gap quantum anomalous Hall states 
in two-dimensional materials with the chemical functionalization, 
and provides opportunities in designing low-power and noise-tolerant spintronic devices.
\end{abstract}
\maketitle

\section{Introduction}

The quantum anomalous Hall state is a topologically nontrivial state 
characterized by the quantization of the Hall conductance $\sigma_{xy}=Ce^{2}/h$ with nonzero Chern number $C$
and chiral edge states, even without the external magnetic field \cite{Haldane1988,Chang2023}. 
This novel quantum state of matter has been experimentally realized
in magnetically doped topological insulators \cite{Chang2013, Chang2015, Mogi2015}, 
intrinsic magnetic topological insulators \cite{Deng2020}, 
and moir\'e materials  \cite{Serlin2020,li2021}.
However, the experimental temperature of realizing the nontrivial state is still low, 
which brings difficulties to observations and applications. 
A large nontrivial band gap is thus expected to immune to the influence of temperature, disorder, etc, 
and increases the experimental temperature \cite{Chang2016}.
On the other hand, a high Chern number ($|C|>1$) is desirable to provide multiple topological transport channels. 
Therefore, the search for the Chern insulator with a large band gap and a high Chern number 
is vital for practical applications of the quantum anomalous Hall state.

%Fluorination

The chemical functionalization is a widely used method to modulate various properties of two-dimensional (2D) materials  \cite{Kuila2012, Martin2020}. 
For examples, 
the hydrogen functionalizations of graphene and germanane induce a phase transition from a conductive semimetal to an insulator \cite{Elias2009, LiR2021}; 
the hydrogen and fluorine functionalizations of the stanene monolayer and III-Bi bilayers lead to the quantum spin Hall state with large nontrivial band gaps \cite{Xu2013, Ma2015}; 
the fluorine functionalization also introduces the magnetism in graphene, BN and MoS$_{2}$ \cite{ Feng2013, Zhang2009, Gao2015}. 
Recently, one of the authors proposed that the fully fluorinated 1T-MoX$_{2}$ (X=S, Se) monolayer has a stable ferromagnetic order 
with a high Cuire temperature \cite{Wu2023}.

In this paper, by first-principles calculations, 
we study electronic and topological properties of the fully fluorinated 1T-MoSe$_{2}$ monolayer, which is abbreviated as MoSe$_{2}$F$_{2}$ hereafter.  
The MoSe$_{2}$F$_{2}$ monolayer is a half semi-metal with the trigonal warping around $K_{\pm}$ valleys in its electronic band structure,
 when the spin-orbit coupling is absent. 
Remarkably, the spin-orbit coupling gives rise to large band gaps of $117.2$ meV near two valleys.
By calculating the Chern number and edge states, 
the quantum anomalous Hall state with $|C|=2$ is verified in the gapped monolayer. 
Therefore, a Chern insulator proposed here has a large band gap and multiple conducting channels, 
which are likely to make the MoSe$_{2}$F$_{2}$ monolayer standout among the candidates of the quantum anomalous Hall state. 
Last but not least, 
an effective $\textbf{\emph{k}}\cdot \textbf{\emph{p}}$ model is provided, 
and it captures the origin of the band topology of this monolayer. 
Our work broadens the material choices of the quantum anomalous Hall state, 
and offers potential applications in designing energy-efficient spintronic devices.

\section{Methods}

The first-principles calculations within the framework of the density functional theory \cite{Hohenberg1964, Kohn1965}
 are performed to investigate electronic properties of the MoSe$_{2}$F$_{2}$ monolayer, 
 as implemented in Vienna \textit{Ab} \textit{initio} Simulation Package \cite{Kresse1996a, Kresse1996b}. 
  The projected augmented wave method  \cite{Blochl1994} 
and the Perdew-Burke-Ernzerhof exchange-correlation functionals  \cite{Perdew1996} are adopted,
 with an energy cutoff of 550 eV. 
A $\Gamma$-centered $k$-mesh  \cite{Monkhorst1976} of $10\times10\times1$  is used to sample in the first Brillouin zone of the 2D monolayer.
 A vacuum layer of more than 17 \AA \space is inserted to avoid the interaction between the monolayer and its periodic images, 
The convergence thresholds of the total energy and the interatomic force on each atom 
are set to 10$^{-6}$ eV and $10^{-3}$ eV/\AA , respectively. 
 To consider the correlation effects of 3$d$ electrons of Mo atom, 
 the onsite Coulomb corrections  \cite{Dudarev1998}
 are applied with an effective $U$ of 3 eV, 
 which gives a similar nonrelativisitic band structure with the HSE06 method \cite{Heyd2003, Wu2023}. 
 The magnified band structures at $K_{\pm}$ valleys are obtained by Wannier interpolation via the Wannier90 package \cite{Pizzi2020}, 
 which is also used to calculate the Berry curvature and anomalous Hall conductivity.

\section{Results}

\subsection{Crystal structure of the MoSe$_{2}$F$_{2}$ monolayer}

\begin{figure}[htb]
\includegraphics[width=0.99\linewidth]{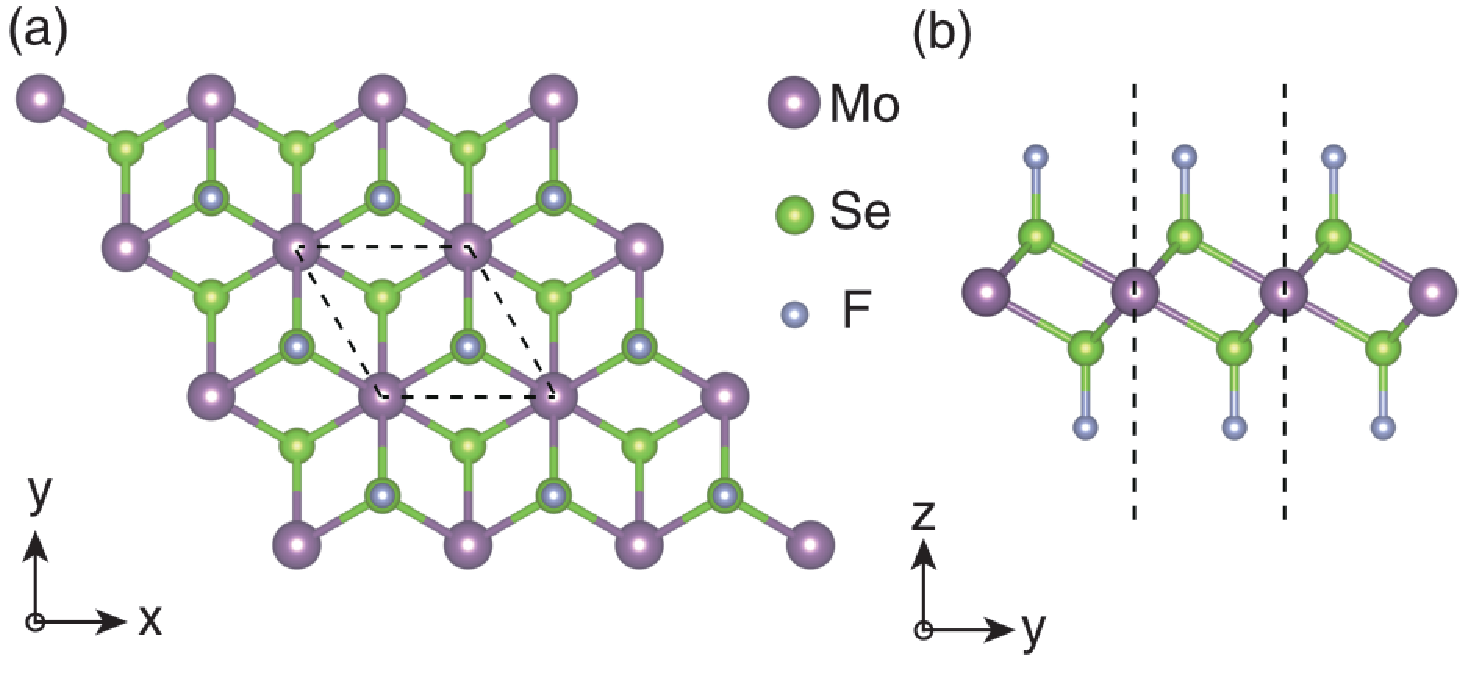}
\caption{Crystal structure of the MoSe$_{2}$F$_{2}$ monolayer. 
(a) The top view and (b) the side view. 
Purple, green and grey balls stand for Mo, Se and F atoms, respectively. 
The unit cell is bounded by dashed lines, exhibiting a hexagonal lattice.}
\label{fig:crystal}
\end{figure}

The crystal structure of the MoSe$_{2}$F$_{2}$ monolayer is shown in Fig. \ref{fig:crystal}.
The MoSe$_{2}$F$_{2}$ monolayer has a point group of $D_{3d}$,
 with the spatial inversion symmetry and three-fold rotational symmetry. 
The monolayer exhibits a hexagonal lattice and comprises five atomic layers. 
The middle three atomic layers are MoSe$_{2}$ in the T phase, 
i.e. each molybdenum atom is bonded to six neighboring selenium atoms that form an octahedral environment. 
The added fluorine atoms reside on the right top (bottom) of the upper (lower) selenium atomic layer. 
For the MoSe$_{2}$F$_{2}$ monolayer, the in-plane lattice constant of the unit cell, $a$, is computed to be  3.94 \AA. 
The distance between the  Se and F  atoms is 1.82 \AA, 
which is smaller than the sum of the covalent radii of Se (1.16 \AA) and F (0.71 \AA), 
indicating the chemical bondings between the Se and absorbed F atoms.
The calculation results are consistent with the previous report about the MoSe$_{2}$F$_{2}$ monolayer, 
where its dynamical and thermodynamical stabilities have also been verified by the calculated phonon spectrum and  $ab$ $initio$ molecular dynamics, respectively \cite{Wu2023}. 
Besides, we calculate the magnetic moment of the ferromagnetic monolayer, 
which has a magnitude of $2.04$ $\mu_{\text{B}}$ per unit cell and agrees with the previous work \cite{Wu2023}.

%===================================================
%===================================================

\subsection{Band structures of the MoSe$_{2}$F$_{2}$ monolayer}

In the followings,  
we investigate electronic properties of the MoSe$_{2}$F$_{2}$ monolayer, without and with the spin-orbit coupling. 
Fig. \ref{eq:2band}(a) shows the band structure without the spin-orbit coupling. 
Owing to the presence of the ferromagnetic order, 
all bands are spin polarized.
While the spin-up channel is semiconducting with a band gap of  $1.41$ eV,
the spin-down channel presents a gapless semi-metallic feature
 with quadratic band touching regions around the Fermi level. 
The band touching regions are located in the neighborhood of the high-symmetry $K_{+}$ and $K_{-}$ points of the momentum space. 
Therefore, the monolayer is a half semi-metal \cite{Cai2015}.

We then focus on the band touching regions around $K_{\pm}$ valleys. 
Given that the monolayer has the spatial inversion symmetry and 
 correspondingly electronic states at $K_{\pm}$ valleys are degenerate, 
we only present the magnified band structures at $K_{+}$ valley. 
The magnified band structure is provided along high-symmetry paths in Fig. \ref{eq:2band}(b). 
It is seen that the valence and conduction bands indeed have quadratic dispersions.
Between the two bands, there are two crossing points, 
which are located at $K_{+}$ point and an intermediate point in the $K_{+}$-$\Gamma$ path, respectively.
For convenience, the wave vector of the crossing point in the  $K_{+}$-$\Gamma$ path is labelled as $S$. 
The energies of the two crossing points are unequal, 
with the one at $K_{+}$ being higher than that at $S$ by $2.5$ meV. 

Fig. \ref{eq:2band}(c) further presents the energy difference between the valence and conduction bands 
over a small 2D region centered at $K_{+}$. 
The energy difference is exactly zero at $K_{+}$ and $S$. 
Besides, owing to the presence of the three-fold rotational symmetry, 
there are three equivalent $S$ points along three equivalent $K_{+}$-$\Gamma$ paths, respectively.
That is, besides the central crossing point at $K_{+}$, 
there are three satellite crossing points with an angular spacing of $2\pi/3$ around $K_{+}$. 
Moreover, the 3D view of the band structure and the energy contours  
 are given around the $K_{+}$ valley for both the valence and conduction bands in Supporting Information (SI hereafter), 
 and they support our findings above. 
According to the energy difference in Fig. \ref{eq:2band}(c) and energy contours in SI, 
it is also found that there is a change of the topology of the Fermi surface as the shift of the chemical potential.
When moving away from the crossing points, 
the Fermi surface first includes four pockets centered at $K_{+}$ and three $S$ points, respectively, 
and then becomes a distorted ring with the trigonal warping. 
This exotic band structure indicates that
the MoSe$_{2}$F$_{2}$ monolayer is a potential candidate to investigate the Lifshitz transition of the Fermi surface, 
by modulating the chemical potential via the carrier doping \cite{Volovik2013, Volovik2017}.

%===================================================
\begin{figure}[htb]
\includegraphics[width=0.99\linewidth]{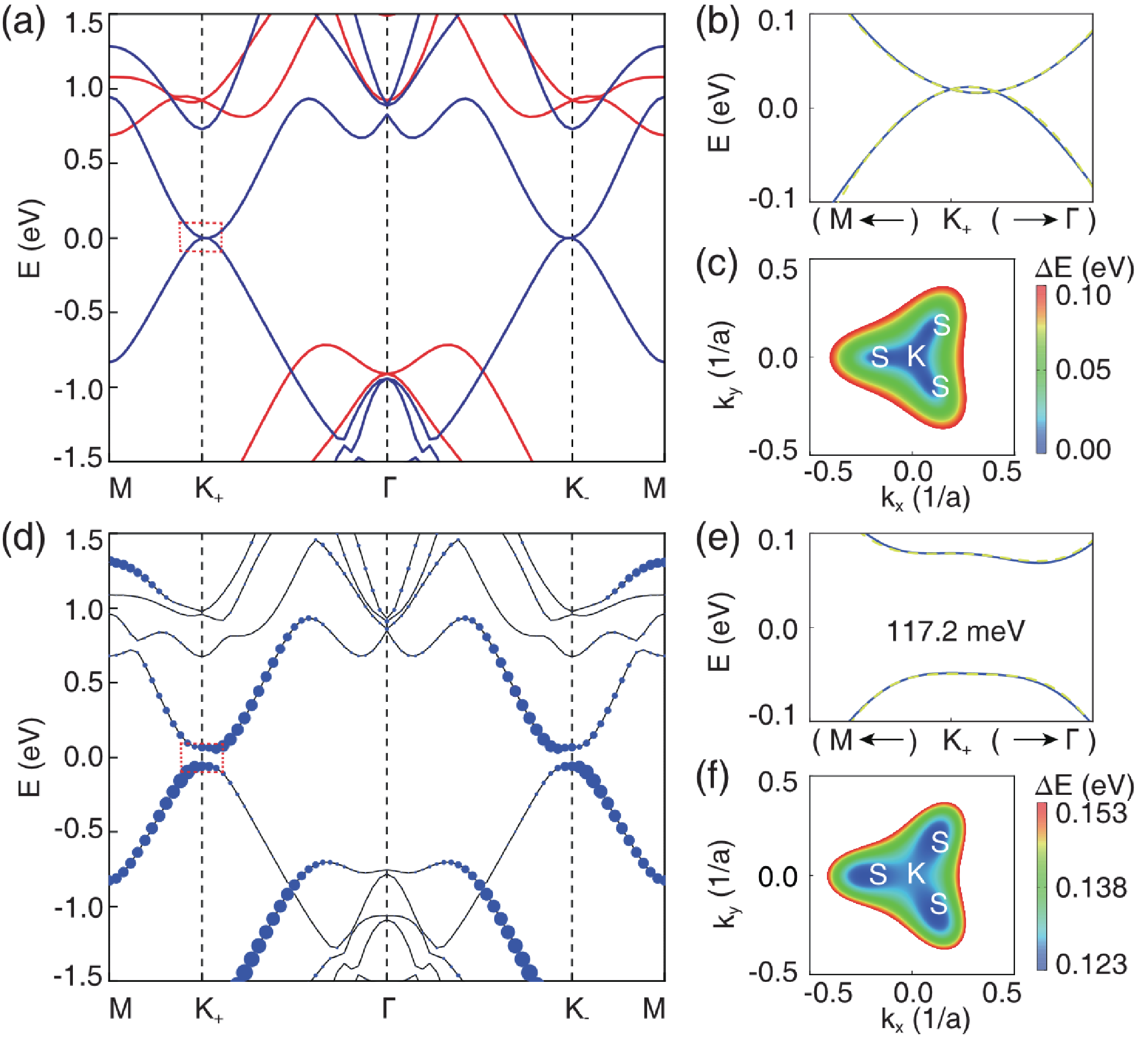}
\caption{Electronic band structures of the MoSe$_{2}$F$_{2}$ monolayer. 
(a) Non-relativistic band structure with spin polarization. 
The spin-up and spin-down channels are represented by red and blue lines, respectively. 
(b) The magnified band structure around $K_{+}$ valley, 
with blue solid lines and yellow dashed lines denoting the bands from the first-principles calculation and the  $k \cdot p$ model, respectively.
(c) The momentum-resolved energy difference between the valence and conduction bands near $K_{+}$ valley. 
(d)-(f) The corresponding band structures and the energy difference for the case with the spin-orbit coupling. 
The Fermi energy is set to zero in band structures. 
In (d), the orbital projections from $d_{xy}$ and $d_{x^2-y^2}$ orbitals of the Mo atoms are denoted by blue circles for each electronic state.
Sizes of circles are proportional to the magnitude of atomic projections.
}
\label{fig:band}
\end{figure}
%===================================================

Fig. \ref{eq:2band}(d) presents the band structures of the MoSe$_{2}$F$_{2}$ monolayer with the spin-orbit coupling. 
 The spin-orbit coupling lifts the band touchings between the valence and conduction bands around $K_{\pm}$ valleys.
 Since the conduction band minimum and the valence band maximum are located at different wave vectors, as illustrated in Fig. \ref{eq:2band}(e), 
 a global indirect band gap appears
  with a magnitude of $117.2$ meV. 
 At $K_{\pm}$ points, the direct band gap has a value of $127.8$ meV.
Therefore, the monolayer becomes a semiconductor due to the introduction of the spin-orbit coupling gaps. 
These band gaps are large enough to survive above the room temperature. 
 Further considering this monolayer is predicted to be a ferromagnet 
 with a Curie temperature of more than 500 K by the Monte Carlo simulation \cite{Wu2023}, 
 the magnetic semiconductivity is robust against the ambient temperature. 
 
 The energy difference between the valence and conduction bands is also shown in Fig. \ref{eq:2band}(f) 
 for the band structure with the spin-orbit coupling. 
 The contour of the energy difference exhibits the trigonal warping and the change of the topology of Fermi surface, 
 similar to the above results without the spin-orbit coupling.

\subsection{The quantum anomalous Hall state with $|C|=2$ }

Given that the band gaps open in a single-spin channel according to the above calculation results 
and the band openings are likely to induce the quantum anomalous Hall state, 
we further investigate topological properties of the MoSe$_{2}$F$_{2}$ monolayer.
We first calculate the momentum-resolved Berry curvature along the $z$ axis, 
$\Omega_{z}(\textbf{\emph{k}})$, which is defined as \cite{Thouless1982,Yao2004,Wang2006,Xiao2010}, 
\be
\Omega_{z}(\textbf{\emph{k}}) =-2\hbar^{2} \text{Im}\sum_{\varepsilon_{nk}\le E_{\text{F}}} \sum_{m\ne n} 
\frac{ \langle \psi_{nk} | v_{x} |\psi_{mk} \rangle\langle\psi_{mk} | v_{y}| \psi_{nk} \rangle}
{(\varepsilon_{mk}-\varepsilon_{nk})^{2}} .
\end{equation}
Here, the Berry curvature includes contributions from all occupied bands below the Fermi level, $E_{\text{F}}$.
$\psi_{nk}$ is the spinor Bloch wave function of the $n$-th band at the wave vector $\textbf{\emph{k}}$, 
with the corresponding electronic energy denoted as $\varepsilon_{nk}$.
$v_{x}$ and $v_{y}$ are the velocity operators along the $x$ and $y$ directions, respectively.

\begin{figure}[htb]
\includegraphics[width=0.99\linewidth]{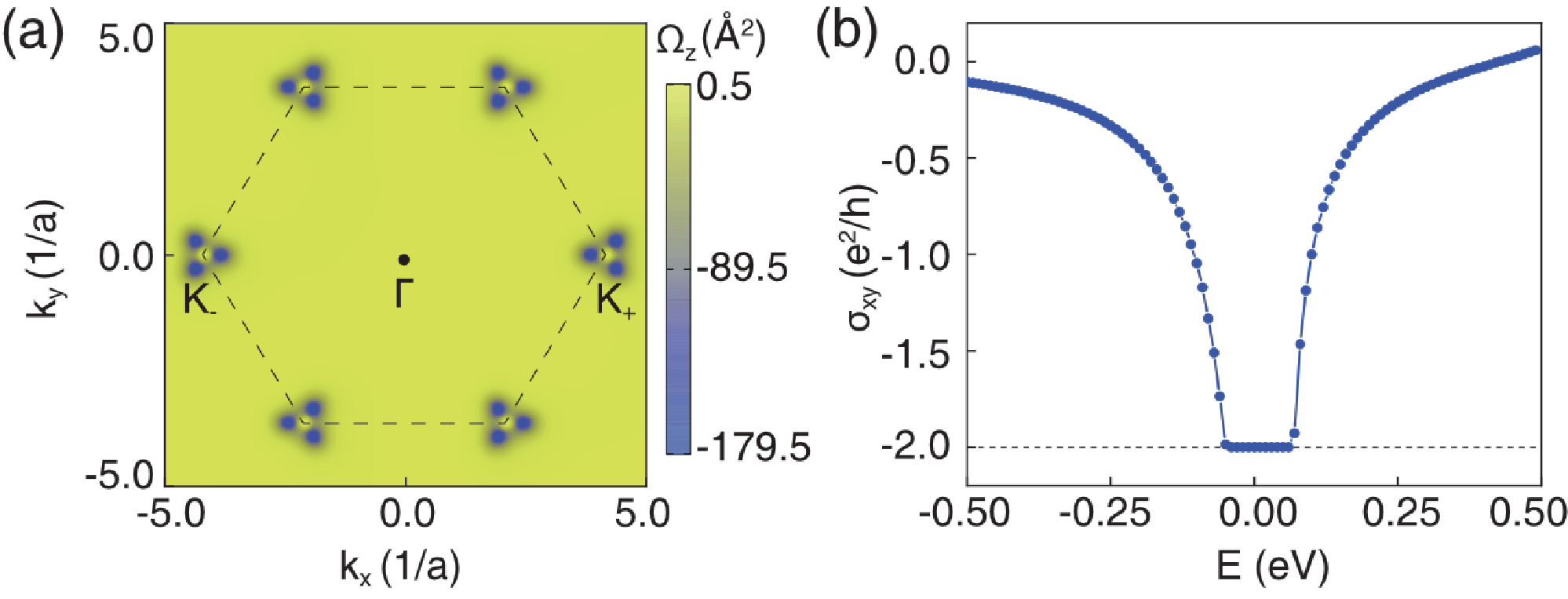}
\caption{Quantum anomalous Hall state of the MoSe$_{2}$F$_{2}$ monolayer. 
 (a) The momentum-resolved Berry curvature. 
(b) The anomalous Hall conductivity as a function of the chemical potential.
}
\label{fig:ahc}
\end{figure}

Fig. \ref{fig:ahc}($a$) demonstrates the Berry curvature in the momentum space.
The Berry curvature is found to mainly concentrate around $K_{+}$ and $K_{-}$ valleys.
Due to the inversion symmetry, the Berry curvatures from the two valleys give the same contribution. 
At each valley, 
the pattern of Berry curvature shows the clover-leaf shape with the three-fold rotational symmetry, 
similar to the energy contour of the band structure. 
It is noted that the pattern is distinct from the single peak of the Berry curvature from massive Dirac cone \cite{Xiao2010}.
By integrating the Berry curvature in the entire first Brillouin zone,
the Chern number of 2D materials can be obtained, 
 i.e. $C =\frac{1}{2\pi} \int_{\text{BZ}} \Omega_{z} (\textbf{\emph{k}}) d^{2} k$  \cite{Thouless1982,Yao2004,Wang2006, Xiao2010}.
Specific to the MoSe$_{2}$F$_{2}$ monolayer,
the calculated Chern number is $C=-2$, 
with the clover-leaf shaped Berry curvature at each valley contributing to a valley Chern number of $-1$. 
The nonzero Chern number indicates the monolayer is indeed a Chern insulator, with a high Chern number. 
As shown in Fig. \ref{fig:ahc}(b), the high Chern number gives rise to a quantized anomalous Hall conductivity. 
A Hall plateau of $-2 e^2/h$ appears, when the chemical potential is within the band gap induced by the spin-orbit coupling.

%===================================================
%===================================================

\begin{figure}[htb]
\includegraphics[width=0.99\linewidth]{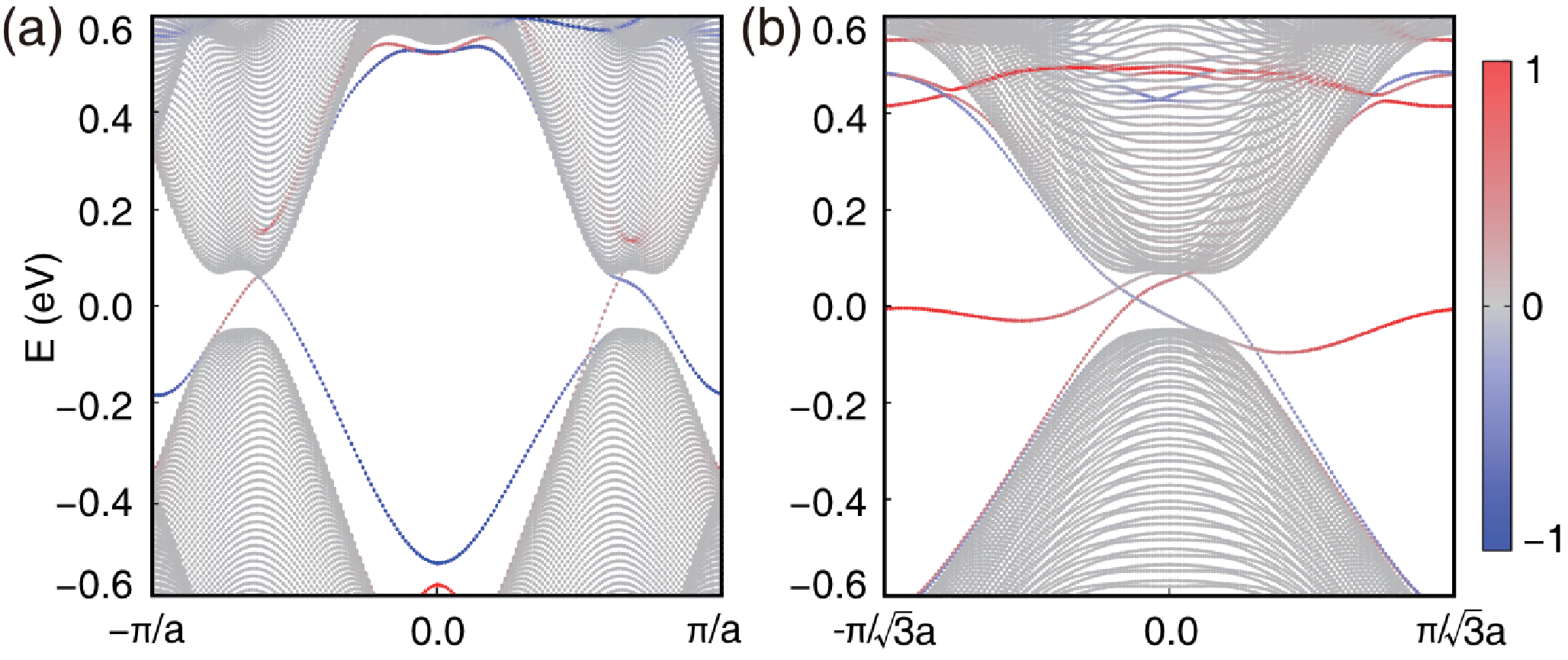}
\caption{Energy spectra of the MoSe$_{2}$F$_{2}$ nanoribbons with (a) zigzag edges and (b) armchair edges, respectively.
The relative distributions at two edges are coded by color.  
The red and blue electronic states denote the ones localized at two opposite edges, respectively, 
while the grey ones correspond to bulk states. 
}
\label{fig:edge}
\end{figure}

Another important indicator for the Chern insulator is 
the existence of topologically protected chiral edge states. 
To further confirm the nontrivial topological properties of the Chern insulator with $C=-2$, 
the energy spectra of the MoSe$_{2}$F$_{2}$ nanoribbons
with the zigzag and armchair edges are calculated by the WannierTools package \cite{Wu2018}, 
and they are presented in Fig. \ref{fig:edge}(a) and  \ref{fig:edge}(b), respectively.
Within the band gap, 
there are topologically nontrivial edge states that seamlessly connect the valence and conduction bands, 
for both the zigzag and armchair nanoribbons. 
These gapless edge states can be divided into two groups that are presented by blue and red lines in Fig. \ref{fig:edge}, respectively. 
Each group includes two topological edge states. 
Topological edge states in different groups have opposite group velocities, 
and they are distributed in opposite edges of the same nanoribbon.  
The topologically nontrivial edge states support the existence of  the quantum anomalous Hall state with $C=-2$.

%===================================================
%===================================================

Moreover, the above results correspond to the magnetization along the $+z$ axis in the MoSe$_{2}$F$_{2}$ monolayer. 
When the magnetization is reversed to the $-z$ axis, 
the Chern number is found to be changed into $+2$. 
Accordingly, the quantized anomalous Hall conductivity becomes $+2 e^2/h$, and the chirality of the edge states is reversed.

\subsection{The low-energy effective model}
\label{sect:trigonal}

To elucidate the origin of the band topology of the MoSe$_{2}$F$_{2}$ monolayer, 
we analyze orbital components of low-energy bands near the Fermi level. 
It is found that the valence and conduction band edges 
are mainly contributed by $d_{xy}$ and $d_{x^2-y^2}$ orbitals of Mo atoms, as shown in Fig. \ref{fig:band}(d). 
Besides, there are also contributions from $p_{z}$ orbitals of Se and F atoms, which is provided in SI. 
We then construct the low-energy $\textbf{\emph{k}}\cdot \textbf{\emph{p}}$ model at $K_{\pm}$ valleys of the monolayer via the MagneticTB package \cite{Zhang2022}, 
using only the $d_{-2}$ (i.e. $d_{x^{2}-y^{2}}/\sqrt{2}-i d_{xy}/\sqrt{2}$) and $d_{+2}$  (i.e. $d_{x^{2}-y^{2}}/\sqrt{2}+i d_{xy}/\sqrt{2}$) orbitals of Mo atoms for simplicity. 
The low-energy model reads, 
\bea
H(\textbf{\emph{q}})= (\varepsilon_{0}-\lambda_{0}q^{2})\sigma_{0} + [\tau\lambda_{1} q_{x} + \lambda_{2}(q_{x}^{2}-q_{y}^{2}) ] \sigma_{x} \nn \\ 
+ [\tau \lambda_{1} q_{y} -2\lambda_{2}q_{x}q_{y}  ]\sigma_{y}+ (\lambda_{3}q^{2} - \frac{\Delta}{2}) \sigma_{z}. 
\label{eq:2band}
\eea
Here, $\textbf{\emph{q}}=(q_{x},q_{y})$ is the wave vector with respect to $K_{\pm}$, and $q^{2}=q_{x}^{2}+q_{y}^{2}$. 
$\sigma_{x,y,z}$ are Pauli matrices for the orbital space, 
 and $\sigma_{0}$ is the corresponding $2\times2$ identity matrix. 
 $\tau=\pm1$ denote $K_{\pm}$ valleys, respectively. 
 $\varepsilon_{0}$ is an immaterial global energy shift. 
 $\lambda_{0}$-$\lambda_{2}$ are strength parameters related to onsite energy and hoppings. 
 $\lambda_{3}$- and $\Delta$-related terms arise from the spin-orbit coupling, 
 and $\Delta$ is equal to the magnitude of the direct band gap at $K_{\pm}$.
 The low-energy effective model is considered up to the second order in $\textbf{\emph{q}}$. 
 It has a similar form with that of the 2H-MoS$_{2}$ monolayer \cite{Falko2013, Liu2013}, but uses different orbital basis sets.

By fitting the first-principles non-relativistic valence and conduction bands via the least squares method, 
the parameters of the low-energy effective model without considering the spin-orbit coupling are obtained as 
 $\varepsilon_{0}=20.0$ meV, $\lambda_{0}=1283.8$ meV$\cdot$\AA$^{2}$, $\lambda_{1}=298.7$ meV$\cdot$\AA, 
$\lambda_{2}=7114.5$ meV$\cdot$\AA$^{2}$. 
When further considering the spin-orbit coupling, 
$\lambda_{3}$ and $\Delta$ are fitted to be $714.1$ meV$\cdot$\AA$^{2}$ and $128.0$ meV, respectively, 
and $\varepsilon_{0}$ becomes $14.6$ meV, while the other parameters are fixed. 
As shown in Figs. \ref{eq:2band}(b) and \ref{eq:2band}(e), 
with this set of parameters,
the band structures from the low-energy effective model and first-principles calculations 
agree with each other very well. 
Furthermore, the clover-leaf shaped Berry curvature is also well reproduced, 
based on our effective model, which is given in SI.

\section{Discussions}

In the above low-energy model, 
we consider the first-order and second-order terms in $\textbf{\emph{q}}$. 
It is found that the corresponding parameters are all sizable. 
Given that the first-order and second-order terms give rise to linear and quadratic band dispersions, respectively, 
the simultaneous presence of these terms lead to the trigonal warping and 
the Lifshitz transition of the topology of the Fermi surface in the MoSe$_{2}$F$_{2}$ monolayer  \cite{Volovik2013,Volovik2017}.
Correspondingly, the Berry curvature with the clover-leaf shape is also due to these terms. 
Moreover, by tuning the ratio of $\lambda_{1}$ to $\lambda_{2}$, 
the band structure and Berry curvature are changed accordingly. 
In particular, as the ratio is increased, 
the shape of the Berry curvature at each valley is changed from a ring to a clover leaf, 
and to a combination of one positive and three negative peaks, 
which is shown in SI. 
Therefore, a tunable ratio is expected to be a practical knob to modulate the band structure and Berry curvature.
 It is likely to be realized by e.g. the strain \cite{Ribeiro2009}.

The quantum anomalous Hall state in the MoSe$_{2}$F$_{2}$ monolayer is intriguing with outstanding advantages. 
Firstly, the monolayer has a high ferromagnetic Cuire temperature \cite{Wu2023} and a considerable band gap of more than $100$ meV, 
which ensures the robustness of the topological state against the temperature. 
Secondly, the Chern number of $|C|=2$ provides more dissipationless conducting channels from topologically nontrivial edge states. 
Moreover, the comparable linear and quadratic terms in low-energy effective model 
lead to potential Lifshitz transition, exotic Berry curvatures around $K_{\pm}$ valleys, and nontrivial band topology. 
The mechanism of the band topology is different from the previous reports about
 other quantum anomalous Hall insulators \cite{Qiao2010, Xu2011, Fang2012, Wang2013, Cai2015, He2017, Li2017, Yang2019}. 
The MoSe$_{2}$F$_{2}$ monolayer provides a new platform for exploring topological quantum states. 

It is noted that the fluorination is one of the chemical modifications of the 1T-MoSe$_{2}$ monolayer. 
The 1T-MoSe$_{2}$ monolayer with other decorated atoms, 
such as hydrogen and iodine atoms, 
may lead to the ferromagnetic order and the band topology, similar to the MoSe$_{2}$F$_{2}$ monolayer.  
Furthermore, different decorated atoms possibly modulate the linear and quadratic contributions to the low-energy band structure. 
They are worth further studies.

\section{Conclusion}

To summarize, we have employed the first-principles calculations to study 
electronic and topological properties of the MoSe$_{2}$F$_{2}$ monolayer. 
In the absence of the spin-orbit coupling, 
the single-spin valence and conduction bands exhibit half semi-metallic features 
and the trigonal warping around $K_{\pm}$ valleys in electronic band structures.  
The Lifshitz transition can occur by tuning the chemical potential.
More interestingly, 
the spin-orbit coupling can induce a global band gap of 117.2 meV near the $K_{\pm}$ valleys. 
Accordingly, 
the quantum anomalous Hall state appears with $|C|=2$, 
which is supported by the calculations of the anomalous Hall conductivity 
and topological edge states. 
A two-band low-energy effective model is also provided to better understand 
the origin of the band topology. 
The quantum anomalous Hall state in the MoSe$_{2}$F$_{2}$ monolayer 
offers multiple dissipationless transport channels and potential high-temperature applications in spintronics.

%-------------------------------------------
\acknowledgements
We are grateful to Jing Wu for valuable discussions. 
We are supported by the
National Natural Science Foundation of China (Nos. 12004186, 12374044, 11904173). 
Xiao Li is also supported by the Jiangsu Specially-Appointed Professor Program. 
%-------------------------------------------



\begin{thebibliography}{48}%
\makeatletter
\providecommand \@ifxundefined [1]{%
 \@ifx{#1\undefined}
}%
\providecommand \@ifnum [1]{%
 \ifnum #1\expandafter \@firstoftwo
 \else \expandafter \@secondoftwo
 \fi
}%
\providecommand \@ifx [1]{%
 \ifx #1\expandafter \@firstoftwo
 \else \expandafter \@secondoftwo
 \fi
}%
\providecommand \natexlab [1]{#1}%
\providecommand \enquote  [1]{``#1''}%
\providecommand \bibnamefont  [1]{#1}%
\providecommand \bibfnamefont [1]{#1}%
\providecommand \citenamefont [1]{#1}%
\providecommand \href@noop [0]{\@secondoftwo}%
\providecommand \href [0]{\begingroup \@sanitize@url \@href}%
\providecommand \@href[1]{\@@startlink{#1}\@@href}%
\providecommand \@@href[1]{\endgroup#1\@@endlink}%
\providecommand \@sanitize@url [0]{\catcode `\\12\catcode `\$12\catcode `\&12\catcode `\#12\catcode `\^12\catcode `\_12\catcode `\%12\relax}%
\providecommand \@@startlink[1]{}%
\providecommand \@@endlink[0]{}%
\providecommand \url  [0]{\begingroup\@sanitize@url \@url }%
\providecommand \@url [1]{\endgroup\@href {#1}{\urlprefix }}%
\providecommand \urlprefix  [0]{URL }%
\providecommand \Eprint [0]{\href }%
\providecommand \doibase [0]{http://dx.doi.org/}%
\providecommand \selectlanguage [0]{\@gobble}%
\providecommand \bibinfo  [0]{\@secondoftwo}%
\providecommand \bibfield  [0]{\@secondoftwo}%
\providecommand \translation [1]{[#1]}%
\providecommand \BibitemOpen [0]{}%
\providecommand \bibitemStop [0]{}%
\providecommand \bibitemNoStop [0]{.\EOS\space}%
\providecommand \EOS [0]{\spacefactor3000\relax}%
\providecommand \BibitemShut  [1]{\csname bibitem#1\endcsname}%
\let\auto@bib@innerbib\@empty
%</preamble>
\bibitem [{\citenamefont {Haldane}(1988)}]{Haldane1988}%
  \BibitemOpen
  \bibfield  {author} {\bibinfo {author} {\bibfnamefont {F.~D.~M.}\ \bibnamefont {Haldane}},\ }\href {\doibase 10.1103/PhysRevLett.61.2015} {\bibfield  {journal} {\bibinfo  {journal} {Phys. Rev. Lett.}\ }\textbf {\bibinfo {volume} {61}},\ \bibinfo {pages} {2015} (\bibinfo {year} {1988})}\BibitemShut {NoStop}%
\bibitem [{\citenamefont {Chang}\ \emph {et~al.}(2023)\citenamefont {Chang}, \citenamefont {Liu},\ and\ \citenamefont {MacDonald}}]{Chang2023}%
  \BibitemOpen
  \bibfield  {author} {\bibinfo {author} {\bibfnamefont {C.-Z.}\ \bibnamefont {Chang}}, \bibinfo {author} {\bibfnamefont {C.-X.}\ \bibnamefont {Liu}}, \ and\ \bibinfo {author} {\bibfnamefont {A.~H.}\ \bibnamefont {MacDonald}},\ }\href {\doibase 10.1103/RevModPhys.95.011002} {\bibfield  {journal} {\bibinfo  {journal} {Rev. Mod. Phys.}\ }\textbf {\bibinfo {volume} {95}},\ \bibinfo {pages} {011002} (\bibinfo {year} {2023})}\BibitemShut {NoStop}%
\bibitem [{\citenamefont {Chang}\ \emph {et~al.}(2013)\citenamefont {Chang}, \citenamefont {Zhang}, \citenamefont {Feng}, \citenamefont {Shen}, \citenamefont {Zhang}, \citenamefont {Guo}, \citenamefont {Li}, \citenamefont {Ou}, \citenamefont {Wei}, \citenamefont {Wang}, \citenamefont {Ji}, \citenamefont {Feng}, \citenamefont {Ji}, \citenamefont {Chen}, \citenamefont {Jia}, \citenamefont {Dai}, \citenamefont {Fang}, \citenamefont {Zhang}, \citenamefont {He}, \citenamefont {Wang}, \citenamefont {Lu}, \citenamefont {Ma},\ and\ \citenamefont {Xue}}]{Chang2013}%
  \BibitemOpen
  \bibfield  {author} {\bibinfo {author} {\bibfnamefont {C.-Z.}\ \bibnamefont {Chang}}, \bibinfo {author} {\bibfnamefont {J.}~\bibnamefont {Zhang}}, \bibinfo {author} {\bibfnamefont {X.}~\bibnamefont {Feng}}, \bibinfo {author} {\bibfnamefont {J.}~\bibnamefont {Shen}}, \bibinfo {author} {\bibfnamefont {Z.}~\bibnamefont {Zhang}}, \bibinfo {author} {\bibfnamefont {M.}~\bibnamefont {Guo}}, \bibinfo {author} {\bibfnamefont {K.}~\bibnamefont {Li}}, \bibinfo {author} {\bibfnamefont {Y.}~\bibnamefont {Ou}}, \bibinfo {author} {\bibfnamefont {P.}~\bibnamefont {Wei}}, \bibinfo {author} {\bibfnamefont {L.-L.}\ \bibnamefont {Wang}}, \bibinfo {author} {\bibfnamefont {Z.-Q.}\ \bibnamefont {Ji}}, \bibinfo {author} {\bibfnamefont {Y.}~\bibnamefont {Feng}}, \bibinfo {author} {\bibfnamefont {S.}~\bibnamefont {Ji}}, \bibinfo {author} {\bibfnamefont {X.}~\bibnamefont {Chen}}, \bibinfo {author} {\bibfnamefont {J.}~\bibnamefont {Jia}}, \bibinfo {author} {\bibfnamefont {X.}~\bibnamefont {Dai}}, \bibinfo {author} {\bibfnamefont
  {Z.}~\bibnamefont {Fang}}, \bibinfo {author} {\bibfnamefont {S.-C.}\ \bibnamefont {Zhang}}, \bibinfo {author} {\bibfnamefont {K.}~\bibnamefont {He}}, \bibinfo {author} {\bibfnamefont {Y.}~\bibnamefont {Wang}}, \bibinfo {author} {\bibfnamefont {L.}~\bibnamefont {Lu}}, \bibinfo {author} {\bibfnamefont {X.-C.}\ \bibnamefont {Ma}}, \ and\ \bibinfo {author} {\bibfnamefont {Q.-K.}\ \bibnamefont {Xue}},\ }\href {\doibase 10.1126/science.1234414} {\bibfield  {journal} {\bibinfo  {journal} {Science}\ }\textbf {\bibinfo {volume} {340}},\ \bibinfo {pages} {167} (\bibinfo {year} {2013})}\BibitemShut {NoStop}%
\bibitem [{\citenamefont {Chang}\ \emph {et~al.}(2015)\citenamefont {Chang}, \citenamefont {Zhao}, \citenamefont {Kim}, \citenamefont {Zhang}, \citenamefont {Assaf}, \citenamefont {Heiman}, \citenamefont {Zhang}, \citenamefont {Liu}, \citenamefont {Chan},\ and\ \citenamefont {Moodera}}]{Chang2015}%
  \BibitemOpen
  \bibfield  {author} {\bibinfo {author} {\bibfnamefont {C.-Z.}\ \bibnamefont {Chang}}, \bibinfo {author} {\bibfnamefont {W.}~\bibnamefont {Zhao}}, \bibinfo {author} {\bibfnamefont {D.~Y.}\ \bibnamefont {Kim}}, \bibinfo {author} {\bibfnamefont {H.}~\bibnamefont {Zhang}}, \bibinfo {author} {\bibfnamefont {B.~A.}\ \bibnamefont {Assaf}}, \bibinfo {author} {\bibfnamefont {D.}~\bibnamefont {Heiman}}, \bibinfo {author} {\bibfnamefont {S.-C.}\ \bibnamefont {Zhang}}, \bibinfo {author} {\bibfnamefont {C.}~\bibnamefont {Liu}}, \bibinfo {author} {\bibfnamefont {M.~H.~W.}\ \bibnamefont {Chan}}, \ and\ \bibinfo {author} {\bibfnamefont {J.~S.}\ \bibnamefont {Moodera}},\ }\href {\doibase 10.1038/nmat4204} {\bibfield  {journal} {\bibinfo  {journal} {Nature Materials}\ }\textbf {\bibinfo {volume} {14}},\ \bibinfo {pages} {473} (\bibinfo {year} {2015})}\BibitemShut {NoStop}%
\bibitem [{\citenamefont {Mogi}\ \emph {et~al.}(2015)\citenamefont {Mogi}, \citenamefont {Yoshimi}, \citenamefont {Tsukazaki}, \citenamefont {Yasuda}, \citenamefont {Kozuka}, \citenamefont {Takahashi}, \citenamefont {Kawasaki},\ and\ \citenamefont {Tokura}}]{Mogi2015}%
  \BibitemOpen
  \bibfield  {author} {\bibinfo {author} {\bibfnamefont {M.}~\bibnamefont {Mogi}}, \bibinfo {author} {\bibfnamefont {R.}~\bibnamefont {Yoshimi}}, \bibinfo {author} {\bibfnamefont {A.}~\bibnamefont {Tsukazaki}}, \bibinfo {author} {\bibfnamefont {K.}~\bibnamefont {Yasuda}}, \bibinfo {author} {\bibfnamefont {Y.}~\bibnamefont {Kozuka}}, \bibinfo {author} {\bibfnamefont {K.~S.}\ \bibnamefont {Takahashi}}, \bibinfo {author} {\bibfnamefont {M.}~\bibnamefont {Kawasaki}}, \ and\ \bibinfo {author} {\bibfnamefont {Y.}~\bibnamefont {Tokura}},\ }\href {https://doi.org/10.1063/1.4935075} {\bibfield  {journal} {\bibinfo  {journal} {Applied Physics Letters}\ }\textbf {\bibinfo {volume} {107}},\ \bibinfo {pages} {182401} (\bibinfo {year} {2015})}\BibitemShut {NoStop}%
\bibitem [{\citenamefont {Deng}\ \emph {et~al.}(2020)\citenamefont {Deng}, \citenamefont {Yu}, \citenamefont {Shi}, \citenamefont {Guo}, \citenamefont {Xu}, \citenamefont {Wang}, \citenamefont {Chen},\ and\ \citenamefont {Zhang}}]{Deng2020}%
  \BibitemOpen
  \bibfield  {author} {\bibinfo {author} {\bibfnamefont {Y.}~\bibnamefont {Deng}}, \bibinfo {author} {\bibfnamefont {Y.}~\bibnamefont {Yu}}, \bibinfo {author} {\bibfnamefont {M.~Z.}\ \bibnamefont {Shi}}, \bibinfo {author} {\bibfnamefont {Z.}~\bibnamefont {Guo}}, \bibinfo {author} {\bibfnamefont {Z.}~\bibnamefont {Xu}}, \bibinfo {author} {\bibfnamefont {J.}~\bibnamefont {Wang}}, \bibinfo {author} {\bibfnamefont {X.~H.}\ \bibnamefont {Chen}}, \ and\ \bibinfo {author} {\bibfnamefont {Y.}~\bibnamefont {Zhang}},\ }\href {\doibase 10.1126/science.aax8156} {\bibfield  {journal} {\bibinfo  {journal} {Science}\ }\textbf {\bibinfo {volume} {367}},\ \bibinfo {pages} {895} (\bibinfo {year} {2020})}\BibitemShut {NoStop}%
\bibitem [{\citenamefont {Serlin}\ \emph {et~al.}(2020)\citenamefont {Serlin}, \citenamefont {Tschirhart}, \citenamefont {Polshyn}, \citenamefont {Zhang}, \citenamefont {Zhu}, \citenamefont {Watanabe}, \citenamefont {Taniguchi}, \citenamefont {Balents},\ and\ \citenamefont {Young}}]{Serlin2020}%
  \BibitemOpen
  \bibfield  {author} {\bibinfo {author} {\bibfnamefont {M.}~\bibnamefont {Serlin}}, \bibinfo {author} {\bibfnamefont {C.~L.}\ \bibnamefont {Tschirhart}}, \bibinfo {author} {\bibfnamefont {H.}~\bibnamefont {Polshyn}}, \bibinfo {author} {\bibfnamefont {Y.}~\bibnamefont {Zhang}}, \bibinfo {author} {\bibfnamefont {J.}~\bibnamefont {Zhu}}, \bibinfo {author} {\bibfnamefont {K.}~\bibnamefont {Watanabe}}, \bibinfo {author} {\bibfnamefont {T.}~\bibnamefont {Taniguchi}}, \bibinfo {author} {\bibfnamefont {L.}~\bibnamefont {Balents}}, \ and\ \bibinfo {author} {\bibfnamefont {A.~F.}\ \bibnamefont {Young}},\ }\href {\doibase 10.1126/science.aay5533} {\bibfield  {journal} {\bibinfo  {journal} {Science}\ }\textbf {\bibinfo {volume} {367}},\ \bibinfo {pages} {900} (\bibinfo {year} {2020})}\BibitemShut {NoStop}%
\bibitem [{\citenamefont {Li}\ \emph {et~al.}(2021{\natexlab{a}})\citenamefont {Li}, \citenamefont {Jiang}, \citenamefont {Shen}, \citenamefont {Zhang}, \citenamefont {Li}, \citenamefont {Tao}, \citenamefont {Devakul}, \citenamefont {Watanabe}, \citenamefont {Taniguchi}, \citenamefont {Fu}, \citenamefont {Shan},\ and\ \citenamefont {Mak}}]{li2021}%
  \BibitemOpen
  \bibfield  {author} {\bibinfo {author} {\bibfnamefont {T.}~\bibnamefont {Li}}, \bibinfo {author} {\bibfnamefont {S.}~\bibnamefont {Jiang}}, \bibinfo {author} {\bibfnamefont {B.}~\bibnamefont {Shen}}, \bibinfo {author} {\bibfnamefont {Y.}~\bibnamefont {Zhang}}, \bibinfo {author} {\bibfnamefont {L.}~\bibnamefont {Li}}, \bibinfo {author} {\bibfnamefont {Z.}~\bibnamefont {Tao}}, \bibinfo {author} {\bibfnamefont {T.}~\bibnamefont {Devakul}}, \bibinfo {author} {\bibfnamefont {K.}~\bibnamefont {Watanabe}}, \bibinfo {author} {\bibfnamefont {T.}~\bibnamefont {Taniguchi}}, \bibinfo {author} {\bibfnamefont {L.}~\bibnamefont {Fu}}, \bibinfo {author} {\bibfnamefont {J.}~\bibnamefont {Shan}}, \ and\ \bibinfo {author} {\bibfnamefont {K.~F.}\ \bibnamefont {Mak}},\ }\href {\doibase 10.1038/s41586-021-04171-1} {\bibfield  {journal} {\bibinfo  {journal} {Nature}\ }\textbf {\bibinfo {volume} {600}},\ \bibinfo {pages} {641} (\bibinfo {year} {2021}{\natexlab{a}})}\BibitemShut {NoStop}%
\bibitem [{\citenamefont {Chang}\ \emph {et~al.}(2016)\citenamefont {Chang}, \citenamefont {Zhao}, \citenamefont {Li}, \citenamefont {Jain}, \citenamefont {Liu}, \citenamefont {Moodera},\ and\ \citenamefont {Chan}}]{Chang2016}%
  \BibitemOpen
  \bibfield  {author} {\bibinfo {author} {\bibfnamefont {C.-Z.}\ \bibnamefont {Chang}}, \bibinfo {author} {\bibfnamefont {W.}~\bibnamefont {Zhao}}, \bibinfo {author} {\bibfnamefont {J.}~\bibnamefont {Li}}, \bibinfo {author} {\bibfnamefont {J.~K.}\ \bibnamefont {Jain}}, \bibinfo {author} {\bibfnamefont {C.}~\bibnamefont {Liu}}, \bibinfo {author} {\bibfnamefont {J.~S.}\ \bibnamefont {Moodera}}, \ and\ \bibinfo {author} {\bibfnamefont {M.~H.~W.}\ \bibnamefont {Chan}},\ }\href {\doibase 10.1103/PhysRevLett.117.126802} {\bibfield  {journal} {\bibinfo  {journal} {Phys. Rev. Lett.}\ }\textbf {\bibinfo {volume} {117}},\ \bibinfo {pages} {126802} (\bibinfo {year} {2016})}\BibitemShut {NoStop}%
\bibitem [{\citenamefont {Kuila}\ \emph {et~al.}(2012)\citenamefont {Kuila}, \citenamefont {Bose}, \citenamefont {Mishra}, \citenamefont {Khanra}, \citenamefont {Kim},\ and\ \citenamefont {Lee}}]{Kuila2012}%
  \BibitemOpen
  \bibfield  {author} {\bibinfo {author} {\bibfnamefont {T.}~\bibnamefont {Kuila}}, \bibinfo {author} {\bibfnamefont {S.}~\bibnamefont {Bose}}, \bibinfo {author} {\bibfnamefont {A.~K.}\ \bibnamefont {Mishra}}, \bibinfo {author} {\bibfnamefont {P.}~\bibnamefont {Khanra}}, \bibinfo {author} {\bibfnamefont {N.~H.}\ \bibnamefont {Kim}}, \ and\ \bibinfo {author} {\bibfnamefont {J.~H.}\ \bibnamefont {Lee}},\ }\href {\doibase https://doi.org/10.1016/j.pmatsci.2012.03.002} {\bibfield  {journal} {\bibinfo  {journal} {Progress in Materials Science}\ }\textbf {\bibinfo {volume} {57}},\ \bibinfo {pages} {1061} (\bibinfo {year} {2012})}\BibitemShut {NoStop}%
\bibitem [{\citenamefont {Martín}\ \emph {et~al.}(2020)\citenamefont {Martín}, \citenamefont {Tagmatarchis}, \citenamefont {Wang},\ and\ \citenamefont {Zhang}}]{Martin2020}%
  \BibitemOpen
  \bibfield  {author} {\bibinfo {author} {\bibfnamefont {N.}~\bibnamefont {Martín}}, \bibinfo {author} {\bibfnamefont {N.}~\bibnamefont {Tagmatarchis}}, \bibinfo {author} {\bibfnamefont {Q.~H.}\ \bibnamefont {Wang}}, \ and\ \bibinfo {author} {\bibfnamefont {X.}~\bibnamefont {Zhang}},\ }\href {\doibase https://doi.org/10.1002/chem.202001304} {\bibfield  {journal} {\bibinfo  {journal} {Chemistry – A European Journal}\ }\textbf {\bibinfo {volume} {26}},\ \bibinfo {pages} {6292} (\bibinfo {year} {2020})}\BibitemShut {NoStop}%
\bibitem [{\citenamefont {Elias}\ \emph {et~al.}(2009)\citenamefont {Elias}, \citenamefont {Nair}, \citenamefont {Mohiuddin}, \citenamefont {Morozov}, \citenamefont {Blake}, \citenamefont {Halsall}, \citenamefont {Ferrari}, \citenamefont {Boukhvalov}, \citenamefont {Katsnelson}, \citenamefont {Geim},\ and\ \citenamefont {Novoselov}}]{Elias2009}%
  \BibitemOpen
  \bibfield  {author} {\bibinfo {author} {\bibfnamefont {D.~C.}\ \bibnamefont {Elias}}, \bibinfo {author} {\bibfnamefont {R.~R.}\ \bibnamefont {Nair}}, \bibinfo {author} {\bibfnamefont {T.~M.~G.}\ \bibnamefont {Mohiuddin}}, \bibinfo {author} {\bibfnamefont {S.~V.}\ \bibnamefont {Morozov}}, \bibinfo {author} {\bibfnamefont {P.}~\bibnamefont {Blake}}, \bibinfo {author} {\bibfnamefont {M.~P.}\ \bibnamefont {Halsall}}, \bibinfo {author} {\bibfnamefont {A.~C.}\ \bibnamefont {Ferrari}}, \bibinfo {author} {\bibfnamefont {D.~W.}\ \bibnamefont {Boukhvalov}}, \bibinfo {author} {\bibfnamefont {M.~I.}\ \bibnamefont {Katsnelson}}, \bibinfo {author} {\bibfnamefont {A.~K.}\ \bibnamefont {Geim}}, \ and\ \bibinfo {author} {\bibfnamefont {K.~S.}\ \bibnamefont {Novoselov}},\ }\href {\doibase 10.1126/science.1167130} {\bibfield  {journal} {\bibinfo  {journal} {Science}\ }\textbf {\bibinfo {volume} {323}},\ \bibinfo {pages} {610} (\bibinfo {year} {2009})}\BibitemShut {NoStop}%
\bibitem [{\citenamefont {Li}\ \emph {et~al.}(2021{\natexlab{b}})\citenamefont {Li}, \citenamefont {Li}, \citenamefont {Yang}, \citenamefont {Huang}, \citenamefont {Zhang}, \citenamefont {Tian}, \citenamefont {Huang}, \citenamefont {Yao}, \citenamefont {Liao}, \citenamefont {Yu}, \citenamefont {Liu}, \citenamefont {Li}, \citenamefont {Huang}, \citenamefont {Guo}, \citenamefont {Mei}, \citenamefont {Wang}, \citenamefont {Li},\ and\ \citenamefont {Liu}}]{LiR2021}%
  \BibitemOpen
  \bibfield  {author} {\bibinfo {author} {\bibfnamefont {R.}~\bibnamefont {Li}}, \bibinfo {author} {\bibfnamefont {Y.}~\bibnamefont {Li}}, \bibinfo {author} {\bibfnamefont {Y.}~\bibnamefont {Yang}}, \bibinfo {author} {\bibfnamefont {X.}~\bibnamefont {Huang}}, \bibinfo {author} {\bibfnamefont {S.}~\bibnamefont {Zhang}}, \bibinfo {author} {\bibfnamefont {H.}~\bibnamefont {Tian}}, \bibinfo {author} {\bibfnamefont {X.}~\bibnamefont {Huang}}, \bibinfo {author} {\bibfnamefont {Z.}~\bibnamefont {Yao}}, \bibinfo {author} {\bibfnamefont {P.}~\bibnamefont {Liao}}, \bibinfo {author} {\bibfnamefont {S.}~\bibnamefont {Yu}}, \bibinfo {author} {\bibfnamefont {S.}~\bibnamefont {Liu}}, \bibinfo {author} {\bibfnamefont {Z.}~\bibnamefont {Li}}, \bibinfo {author} {\bibfnamefont {Y.}~\bibnamefont {Huang}}, \bibinfo {author} {\bibfnamefont {J.}~\bibnamefont {Guo}}, \bibinfo {author} {\bibfnamefont {F.}~\bibnamefont {Mei}}, \bibinfo {author} {\bibfnamefont {L.}~\bibnamefont {Wang}}, \bibinfo {author} {\bibfnamefont
  {X.}~\bibnamefont {Li}}, \ and\ \bibinfo {author} {\bibfnamefont {L.}~\bibnamefont {Liu}},\ }\href@noop {} {\bibfield  {journal} {\bibinfo  {journal} {ACS Applied Nano Materials}\ }\textbf {\bibinfo {volume} {4}},\ \bibinfo {pages} {13708} (\bibinfo {year} {2021}{\natexlab{b}})}\BibitemShut {NoStop}%
\bibitem [{\citenamefont {Xu}\ \emph {et~al.}(2013)\citenamefont {Xu}, \citenamefont {Yan}, \citenamefont {Zhang}, \citenamefont {Wang}, \citenamefont {Xu}, \citenamefont {Tang}, \citenamefont {Duan},\ and\ \citenamefont {Zhang}}]{Xu2013}%
  \BibitemOpen
  \bibfield  {author} {\bibinfo {author} {\bibfnamefont {Y.}~\bibnamefont {Xu}}, \bibinfo {author} {\bibfnamefont {B.}~\bibnamefont {Yan}}, \bibinfo {author} {\bibfnamefont {H.-J.}\ \bibnamefont {Zhang}}, \bibinfo {author} {\bibfnamefont {J.}~\bibnamefont {Wang}}, \bibinfo {author} {\bibfnamefont {G.}~\bibnamefont {Xu}}, \bibinfo {author} {\bibfnamefont {P.}~\bibnamefont {Tang}}, \bibinfo {author} {\bibfnamefont {W.}~\bibnamefont {Duan}}, \ and\ \bibinfo {author} {\bibfnamefont {S.-C.}\ \bibnamefont {Zhang}},\ }\href {\doibase 10.1103/PhysRevLett.111.136804} {\bibfield  {journal} {\bibinfo  {journal} {Phys. Rev. Lett.}\ }\textbf {\bibinfo {volume} {111}},\ \bibinfo {pages} {136804} (\bibinfo {year} {2013})}\BibitemShut {NoStop}%
\bibitem [{\citenamefont {Ma}\ \emph {et~al.}(2015)\citenamefont {Ma}, \citenamefont {Li}, \citenamefont {Kou}, \citenamefont {Yan}, \citenamefont {Niu}, \citenamefont {Dai},\ and\ \citenamefont {Heine}}]{Ma2015}%
  \BibitemOpen
  \bibfield  {author} {\bibinfo {author} {\bibfnamefont {Y.}~\bibnamefont {Ma}}, \bibinfo {author} {\bibfnamefont {X.}~\bibnamefont {Li}}, \bibinfo {author} {\bibfnamefont {L.}~\bibnamefont {Kou}}, \bibinfo {author} {\bibfnamefont {B.}~\bibnamefont {Yan}}, \bibinfo {author} {\bibfnamefont {C.}~\bibnamefont {Niu}}, \bibinfo {author} {\bibfnamefont {Y.}~\bibnamefont {Dai}}, \ and\ \bibinfo {author} {\bibfnamefont {T.}~\bibnamefont {Heine}},\ }\href {\doibase 10.1103/PhysRevB.91.235306} {\bibfield  {journal} {\bibinfo  {journal} {Phys. Rev. B}\ }\textbf {\bibinfo {volume} {91}},\ \bibinfo {pages} {235306} (\bibinfo {year} {2015})}\BibitemShut {NoStop}%
\bibitem [{\citenamefont {Feng}\ \emph {et~al.}(2013)\citenamefont {Feng}, \citenamefont {Tang}, \citenamefont {Liu}, \citenamefont {Cao}, \citenamefont {Zheng}, \citenamefont {Ren}, \citenamefont {Wan},\ and\ \citenamefont {Du}}]{Feng2013}%
  \BibitemOpen
  \bibfield  {author} {\bibinfo {author} {\bibfnamefont {Q.}~\bibnamefont {Feng}}, \bibinfo {author} {\bibfnamefont {N.}~\bibnamefont {Tang}}, \bibinfo {author} {\bibfnamefont {F.}~\bibnamefont {Liu}}, \bibinfo {author} {\bibfnamefont {Q.}~\bibnamefont {Cao}}, \bibinfo {author} {\bibfnamefont {W.}~\bibnamefont {Zheng}}, \bibinfo {author} {\bibfnamefont {W.}~\bibnamefont {Ren}}, \bibinfo {author} {\bibfnamefont {X.}~\bibnamefont {Wan}}, \ and\ \bibinfo {author} {\bibfnamefont {Y.}~\bibnamefont {Du}},\ }\href {\doibase 10.1021/nn4027905} {\bibfield  {journal} {\bibinfo  {journal} {ACS Nano}\ }\textbf {\bibinfo {volume} {7}},\ \bibinfo {pages} {6729} (\bibinfo {year} {2013})},\ \bibinfo {note} {publisher: American Chemical Society}\BibitemShut {NoStop}%
\bibitem [{\citenamefont {Zhang}\ and\ \citenamefont {Guo}(2009)}]{Zhang2009}%
  \BibitemOpen
  \bibfield  {author} {\bibinfo {author} {\bibfnamefont {Z.}~\bibnamefont {Zhang}}\ and\ \bibinfo {author} {\bibfnamefont {W.}~\bibnamefont {Guo}},\ }\href {\doibase 10.1021/ja901586k} {\bibfield  {journal} {\bibinfo  {journal} {J. Am. Chem. Soc.}\ }\textbf {\bibinfo {volume} {131}},\ \bibinfo {pages} {6874} (\bibinfo {year} {2009})},\ \bibinfo {note} {publisher: American Chemical Society}\BibitemShut {NoStop}%
\bibitem [{\citenamefont {Gao}\ \emph {et~al.}(2015)\citenamefont {Gao}, \citenamefont {Shi}, \citenamefont {Tao}, \citenamefont {Xia},\ and\ \citenamefont {Xue}}]{Gao2015}%
  \BibitemOpen
  \bibfield  {author} {\bibinfo {author} {\bibfnamefont {D.}~\bibnamefont {Gao}}, \bibinfo {author} {\bibfnamefont {S.}~\bibnamefont {Shi}}, \bibinfo {author} {\bibfnamefont {K.}~\bibnamefont {Tao}}, \bibinfo {author} {\bibfnamefont {B.}~\bibnamefont {Xia}}, \ and\ \bibinfo {author} {\bibfnamefont {D.}~\bibnamefont {Xue}},\ }\href {\doibase 10.1039/C5NR00409H} {\bibfield  {journal} {\bibinfo  {journal} {Nanoscale}\ }\textbf {\bibinfo {volume} {7}},\ \bibinfo {pages} {4211} (\bibinfo {year} {2015})}\BibitemShut {NoStop}%
\bibitem [{\citenamefont {Wu}\ \emph {et~al.}()\citenamefont {Wu}, \citenamefont {Guo}, \citenamefont {Wu}, \citenamefont {Li},\ and\ \citenamefont {Wu}}]{Wu2023}%
  \BibitemOpen
  \bibfield  {author} {\bibinfo {author} {\bibfnamefont {J.}~\bibnamefont {Wu}}, \bibinfo {author} {\bibfnamefont {R.}~\bibnamefont {Guo}}, \bibinfo {author} {\bibfnamefont {D.}~\bibnamefont {Wu}}, \bibinfo {author} {\bibfnamefont {X.}~\bibnamefont {Li}}, \ and\ \bibinfo {author} {\bibfnamefont {X.}~\bibnamefont {Wu}},\ }\href@noop {} {\ }\Eprint {http://arxiv.org/abs/arXiv:2310.03995} {arXiv:2310.03995} \BibitemShut {NoStop}%
\bibitem [{\citenamefont {Hohenberg}\ and\ \citenamefont {Kohn}(1964)}]{Hohenberg1964}%
  \BibitemOpen
  \bibfield  {author} {\bibinfo {author} {\bibfnamefont {P.}~\bibnamefont {Hohenberg}}\ and\ \bibinfo {author} {\bibfnamefont {W.}~\bibnamefont {Kohn}},\ }\href {\doibase 10.1103/PhysRev.136.B864} {\bibfield  {journal} {\bibinfo  {journal} {Phys. Rev.}\ }\textbf {\bibinfo {volume} {136}},\ \bibinfo {pages} {B864} (\bibinfo {year} {1964})}\BibitemShut {NoStop}%
\bibitem [{\citenamefont {Kohn}\ and\ \citenamefont {Sham}(1965)}]{Kohn1965}%
  \BibitemOpen
  \bibfield  {author} {\bibinfo {author} {\bibfnamefont {W.}~\bibnamefont {Kohn}}\ and\ \bibinfo {author} {\bibfnamefont {L.~J.}\ \bibnamefont {Sham}},\ }\href {\doibase 10.1103/PhysRev.140.A1133} {\bibfield  {journal} {\bibinfo  {journal} {Phys. Rev.}\ }\textbf {\bibinfo {volume} {140}},\ \bibinfo {pages} {A1133} (\bibinfo {year} {1965})}\BibitemShut {NoStop}%
\bibitem [{\citenamefont {Kresse}\ and\ \citenamefont {Furthm\"uller}(1996)}]{Kresse1996a}%
  \BibitemOpen
  \bibfield  {author} {\bibinfo {author} {\bibfnamefont {G.}~\bibnamefont {Kresse}}\ and\ \bibinfo {author} {\bibfnamefont {J.}~\bibnamefont {Furthm\"uller}},\ }\href {\doibase 10.1103/PhysRevB.54.11169} {\bibfield  {journal} {\bibinfo  {journal} {Phys. Rev. B}\ }\textbf {\bibinfo {volume} {54}},\ \bibinfo {pages} {11169} (\bibinfo {year} {1996})}\BibitemShut {NoStop}%
\bibitem [{\citenamefont {Kresse}\ and\ \citenamefont {Furthmüller}(1996)}]{Kresse1996b}%
  \BibitemOpen
  \bibfield  {author} {\bibinfo {author} {\bibfnamefont {G.}~\bibnamefont {Kresse}}\ and\ \bibinfo {author} {\bibfnamefont {J.}~\bibnamefont {Furthmüller}},\ }\href {\doibase https://doi.org/10.1016/0927-0256(96)00008-0} {\bibfield  {journal} {\bibinfo  {journal} {Computational Materials Science}\ }\textbf {\bibinfo {volume} {6}},\ \bibinfo {pages} {15} (\bibinfo {year} {1996})}\BibitemShut {NoStop}%
\bibitem [{\citenamefont {Bl\"ochl}(1994)}]{Blochl1994}%
  \BibitemOpen
  \bibfield  {author} {\bibinfo {author} {\bibfnamefont {P.~E.}\ \bibnamefont {Bl\"ochl}},\ }\href {\doibase 10.1103/PhysRevB.50.17953} {\bibfield  {journal} {\bibinfo  {journal} {Phys. Rev. B}\ }\textbf {\bibinfo {volume} {50}},\ \bibinfo {pages} {17953} (\bibinfo {year} {1994})}\BibitemShut {NoStop}%
\bibitem [{\citenamefont {Perdew}\ \emph {et~al.}(1996)\citenamefont {Perdew}, \citenamefont {Burke},\ and\ \citenamefont {Ernzerhof}}]{Perdew1996}%
  \BibitemOpen
  \bibfield  {author} {\bibinfo {author} {\bibfnamefont {J.~P.}\ \bibnamefont {Perdew}}, \bibinfo {author} {\bibfnamefont {K.}~\bibnamefont {Burke}}, \ and\ \bibinfo {author} {\bibfnamefont {M.}~\bibnamefont {Ernzerhof}},\ }\href {\doibase 10.1103/PhysRevLett.77.3865} {\bibfield  {journal} {\bibinfo  {journal} {Phys. Rev. Lett.}\ }\textbf {\bibinfo {volume} {77}},\ \bibinfo {pages} {3865} (\bibinfo {year} {1996})}\BibitemShut {NoStop}%
\bibitem [{\citenamefont {Monkhorst}\ and\ \citenamefont {Pack}(1976)}]{Monkhorst1976}%
  \BibitemOpen
  \bibfield  {author} {\bibinfo {author} {\bibfnamefont {H.~J.}\ \bibnamefont {Monkhorst}}\ and\ \bibinfo {author} {\bibfnamefont {J.~D.}\ \bibnamefont {Pack}},\ }\href {\doibase 10.1103/PhysRevB.13.5188} {\bibfield  {journal} {\bibinfo  {journal} {Phys. Rev. B}\ }\textbf {\bibinfo {volume} {13}},\ \bibinfo {pages} {5188} (\bibinfo {year} {1976})}\BibitemShut {NoStop}%
\bibitem [{\citenamefont {Dudarev}\ \emph {et~al.}(1998)\citenamefont {Dudarev}, \citenamefont {Botton}, \citenamefont {Savrasov}, \citenamefont {Humphreys},\ and\ \citenamefont {Sutton}}]{Dudarev1998}%
  \BibitemOpen
  \bibfield  {author} {\bibinfo {author} {\bibfnamefont {S.~L.}\ \bibnamefont {Dudarev}}, \bibinfo {author} {\bibfnamefont {G.~A.}\ \bibnamefont {Botton}}, \bibinfo {author} {\bibfnamefont {S.~Y.}\ \bibnamefont {Savrasov}}, \bibinfo {author} {\bibfnamefont {C.~J.}\ \bibnamefont {Humphreys}}, \ and\ \bibinfo {author} {\bibfnamefont {A.~P.}\ \bibnamefont {Sutton}},\ }\href {\doibase 10.1103/PhysRevB.57.1505} {\bibfield  {journal} {\bibinfo  {journal} {Phys. Rev. B}\ }\textbf {\bibinfo {volume} {57}},\ \bibinfo {pages} {1505} (\bibinfo {year} {1998})}\BibitemShut {NoStop}%
\bibitem [{\citenamefont {Heyd}\ \emph {et~al.}(2003)\citenamefont {Heyd}, \citenamefont {Scuseria},\ and\ \citenamefont {Ernzerhof}}]{Heyd2003}%
  \BibitemOpen
  \bibfield  {author} {\bibinfo {author} {\bibfnamefont {J.}~\bibnamefont {Heyd}}, \bibinfo {author} {\bibfnamefont {G.~E.}\ \bibnamefont {Scuseria}}, \ and\ \bibinfo {author} {\bibfnamefont {M.}~\bibnamefont {Ernzerhof}},\ }\href {\doibase 10.1063/1.1564060} {\bibfield  {journal} {\bibinfo  {journal} {The Journal of Chemical Physics}\ }\textbf {\bibinfo {volume} {118}},\ \bibinfo {pages} {8207} (\bibinfo {year} {2003})}\BibitemShut {NoStop}%
\bibitem [{\citenamefont {Pizzi}\ \emph {et~al.}(2020)\citenamefont {Pizzi}, \citenamefont {Vitale}, \citenamefont {Arita}, \citenamefont {Blügel}, \citenamefont {Freimuth}, \citenamefont {Géranton}, \citenamefont {Gibertini}, \citenamefont {Gresch}, \citenamefont {Johnson}, \citenamefont {Koretsune}, \citenamefont {Ibañez-Azpiroz}, \citenamefont {Lee}, \citenamefont {Lihm}, \citenamefont {Marchand}, \citenamefont {Marrazzo}, \citenamefont {Mokrousov}, \citenamefont {Mustafa}, \citenamefont {Nohara}, \citenamefont {Nomura}, \citenamefont {Paulatto}, \citenamefont {Poncé}, \citenamefont {Ponweiser}, \citenamefont {Qiao}, \citenamefont {Thöle}, \citenamefont {Tsirkin}, \citenamefont {Wierzbowska}, \citenamefont {Marzari}, \citenamefont {Vanderbilt}, \citenamefont {Souza}, \citenamefont {Mostofi},\ and\ \citenamefont {Yates}}]{Pizzi2020}%
  \BibitemOpen
  \bibfield  {author} {\bibinfo {author} {\bibfnamefont {G.}~\bibnamefont {Pizzi}}, \bibinfo {author} {\bibfnamefont {V.}~\bibnamefont {Vitale}}, \bibinfo {author} {\bibfnamefont {R.}~\bibnamefont {Arita}}, \bibinfo {author} {\bibfnamefont {S.}~\bibnamefont {Blügel}}, \bibinfo {author} {\bibfnamefont {F.}~\bibnamefont {Freimuth}}, \bibinfo {author} {\bibfnamefont {G.}~\bibnamefont {Géranton}}, \bibinfo {author} {\bibfnamefont {M.}~\bibnamefont {Gibertini}}, \bibinfo {author} {\bibfnamefont {D.}~\bibnamefont {Gresch}}, \bibinfo {author} {\bibfnamefont {C.}~\bibnamefont {Johnson}}, \bibinfo {author} {\bibfnamefont {T.}~\bibnamefont {Koretsune}}, \bibinfo {author} {\bibfnamefont {J.}~\bibnamefont {Ibañez-Azpiroz}}, \bibinfo {author} {\bibfnamefont {H.}~\bibnamefont {Lee}}, \bibinfo {author} {\bibfnamefont {J.-M.}\ \bibnamefont {Lihm}}, \bibinfo {author} {\bibfnamefont {D.}~\bibnamefont {Marchand}}, \bibinfo {author} {\bibfnamefont {A.}~\bibnamefont {Marrazzo}}, \bibinfo {author} {\bibfnamefont
  {Y.}~\bibnamefont {Mokrousov}}, \bibinfo {author} {\bibfnamefont {J.~I.}\ \bibnamefont {Mustafa}}, \bibinfo {author} {\bibfnamefont {Y.}~\bibnamefont {Nohara}}, \bibinfo {author} {\bibfnamefont {Y.}~\bibnamefont {Nomura}}, \bibinfo {author} {\bibfnamefont {L.}~\bibnamefont {Paulatto}}, \bibinfo {author} {\bibfnamefont {S.}~\bibnamefont {Poncé}}, \bibinfo {author} {\bibfnamefont {T.}~\bibnamefont {Ponweiser}}, \bibinfo {author} {\bibfnamefont {J.}~\bibnamefont {Qiao}}, \bibinfo {author} {\bibfnamefont {F.}~\bibnamefont {Thöle}}, \bibinfo {author} {\bibfnamefont {S.~S.}\ \bibnamefont {Tsirkin}}, \bibinfo {author} {\bibfnamefont {M.}~\bibnamefont {Wierzbowska}}, \bibinfo {author} {\bibfnamefont {N.}~\bibnamefont {Marzari}}, \bibinfo {author} {\bibfnamefont {D.}~\bibnamefont {Vanderbilt}}, \bibinfo {author} {\bibfnamefont {I.}~\bibnamefont {Souza}}, \bibinfo {author} {\bibfnamefont {A.~A.}\ \bibnamefont {Mostofi}}, \ and\ \bibinfo {author} {\bibfnamefont {J.~R.}\ \bibnamefont {Yates}},\ }\href {\doibase
  10.1088/1361-648X/ab51ff} {\bibfield  {journal} {\bibinfo  {journal} {Journal of Physics: Condensed Matter}\ }\textbf {\bibinfo {volume} {32}},\ \bibinfo {pages} {165902} (\bibinfo {year} {2020})}\BibitemShut {NoStop}%
\bibitem [{\citenamefont {Cai}\ \emph {et~al.}(2015)\citenamefont {Cai}, \citenamefont {Li}, \citenamefont {Wang}, \citenamefont {Ju}, \citenamefont {Feng},\ and\ \citenamefont {Gong}}]{Cai2015}%
  \BibitemOpen
  \bibfield  {author} {\bibinfo {author} {\bibfnamefont {T.}~\bibnamefont {Cai}}, \bibinfo {author} {\bibfnamefont {X.}~\bibnamefont {Li}}, \bibinfo {author} {\bibfnamefont {F.}~\bibnamefont {Wang}}, \bibinfo {author} {\bibfnamefont {S.}~\bibnamefont {Ju}}, \bibinfo {author} {\bibfnamefont {J.}~\bibnamefont {Feng}}, \ and\ \bibinfo {author} {\bibfnamefont {C.-D.}\ \bibnamefont {Gong}},\ }\href {\doibase 10.1021/acs.nanolett.5b01791} {\bibfield  {journal} {\bibinfo  {journal} {Nano Lett.}\ }\textbf {\bibinfo {volume} {15}},\ \bibinfo {pages} {6434} (\bibinfo {year} {2015})}\BibitemShut {NoStop}%
\bibitem [{\citenamefont {Volovik}(2013)}]{Volovik2013}%
  \BibitemOpen
  \bibfield  {author} {\bibinfo {author} {\bibfnamefont {G.~E.}\ \bibnamefont {Volovik}},\ }\href@noop {} {\bibfield  {journal} {\bibinfo  {journal} {Lecture Notes in Physics}\ }\textbf {\bibinfo {volume} {870}},\ \bibinfo {pages} {343} (\bibinfo {year} {2013})}\BibitemShut {NoStop}%
\bibitem [{\citenamefont {Volovik}(2017)}]{Volovik2017}%
  \BibitemOpen
  \bibfield  {author} {\bibinfo {author} {\bibfnamefont {G.~E.}\ \bibnamefont {Volovik}},\ }\href {\doibase 10.1063/1.4974185} {\bibfield  {journal} {\bibinfo  {journal} {Low Temperature Physics}\ }\textbf {\bibinfo {volume} {43}},\ \bibinfo {pages} {47} (\bibinfo {year} {2017})}\BibitemShut {NoStop}%
\bibitem [{\citenamefont {Thouless}\ \emph {et~al.}(1982)\citenamefont {Thouless}, \citenamefont {Kohmoto}, \citenamefont {Nightingale},\ and\ \citenamefont {den Nijs}}]{Thouless1982}%
  \BibitemOpen
  \bibfield  {author} {\bibinfo {author} {\bibfnamefont {D.~J.}\ \bibnamefont {Thouless}}, \bibinfo {author} {\bibfnamefont {M.}~\bibnamefont {Kohmoto}}, \bibinfo {author} {\bibfnamefont {M.~P.}\ \bibnamefont {Nightingale}}, \ and\ \bibinfo {author} {\bibfnamefont {M.}~\bibnamefont {den Nijs}},\ }\href {\doibase 10.1103/PhysRevLett.49.405} {\bibfield  {journal} {\bibinfo  {journal} {Phys. Rev. Lett.}\ }\textbf {\bibinfo {volume} {49}},\ \bibinfo {pages} {405} (\bibinfo {year} {1982})}\BibitemShut {NoStop}%
\bibitem [{\citenamefont {Yao}\ \emph {et~al.}(2004)\citenamefont {Yao}, \citenamefont {Kleinman}, \citenamefont {MacDonald}, \citenamefont {Sinova}, \citenamefont {Jungwirth}, \citenamefont {Wang}, \citenamefont {Wang},\ and\ \citenamefont {Niu}}]{Yao2004}%
  \BibitemOpen
  \bibfield  {author} {\bibinfo {author} {\bibfnamefont {Y.}~\bibnamefont {Yao}}, \bibinfo {author} {\bibfnamefont {L.}~\bibnamefont {Kleinman}}, \bibinfo {author} {\bibfnamefont {A.~H.}\ \bibnamefont {MacDonald}}, \bibinfo {author} {\bibfnamefont {J.}~\bibnamefont {Sinova}}, \bibinfo {author} {\bibfnamefont {T.}~\bibnamefont {Jungwirth}}, \bibinfo {author} {\bibfnamefont {D.-s.}\ \bibnamefont {Wang}}, \bibinfo {author} {\bibfnamefont {E.}~\bibnamefont {Wang}}, \ and\ \bibinfo {author} {\bibfnamefont {Q.}~\bibnamefont {Niu}},\ }\href {\doibase 10.1103/PhysRevLett.92.037204} {\bibfield  {journal} {\bibinfo  {journal} {Phys. Rev. Lett.}\ }\textbf {\bibinfo {volume} {92}},\ \bibinfo {pages} {037204} (\bibinfo {year} {2004})}\BibitemShut {NoStop}%
\bibitem [{\citenamefont {Wang}\ \emph {et~al.}(2006)\citenamefont {Wang}, \citenamefont {Yates}, \citenamefont {Souza},\ and\ \citenamefont {Vanderbilt}}]{Wang2006}%
  \BibitemOpen
  \bibfield  {author} {\bibinfo {author} {\bibfnamefont {X.}~\bibnamefont {Wang}}, \bibinfo {author} {\bibfnamefont {J.~R.}\ \bibnamefont {Yates}}, \bibinfo {author} {\bibfnamefont {I.}~\bibnamefont {Souza}}, \ and\ \bibinfo {author} {\bibfnamefont {D.}~\bibnamefont {Vanderbilt}},\ }\href {\doibase 10.1103/PhysRevB.74.195118} {\bibfield  {journal} {\bibinfo  {journal} {Phys. Rev. B}\ }\textbf {\bibinfo {volume} {74}},\ \bibinfo {pages} {195118} (\bibinfo {year} {2006})}\BibitemShut {NoStop}%
\bibitem [{\citenamefont {Xiao}\ \emph {et~al.}(2010)\citenamefont {Xiao}, \citenamefont {Chang},\ and\ \citenamefont {Niu}}]{Xiao2010}%
  \BibitemOpen
  \bibfield  {author} {\bibinfo {author} {\bibfnamefont {D.}~\bibnamefont {Xiao}}, \bibinfo {author} {\bibfnamefont {M.-C.}\ \bibnamefont {Chang}}, \ and\ \bibinfo {author} {\bibfnamefont {Q.}~\bibnamefont {Niu}},\ }\href {\doibase 10.1103/RevModPhys.82.1959} {\bibfield  {journal} {\bibinfo  {journal} {Rev. Mod. Phys.}\ }\textbf {\bibinfo {volume} {82}},\ \bibinfo {pages} {1959} (\bibinfo {year} {2010})}\BibitemShut {NoStop}%
\bibitem [{\citenamefont {Wu}\ \emph {et~al.}(2018)\citenamefont {Wu}, \citenamefont {Zhang}, \citenamefont {Song}, \citenamefont {Troyer},\ and\ \citenamefont {Soluyanov}}]{Wu2018}%
  \BibitemOpen
  \bibfield  {author} {\bibinfo {author} {\bibfnamefont {Q.}~\bibnamefont {Wu}}, \bibinfo {author} {\bibfnamefont {S.}~\bibnamefont {Zhang}}, \bibinfo {author} {\bibfnamefont {H.-F.}\ \bibnamefont {Song}}, \bibinfo {author} {\bibfnamefont {M.}~\bibnamefont {Troyer}}, \ and\ \bibinfo {author} {\bibfnamefont {A.~A.}\ \bibnamefont {Soluyanov}},\ }\href {\doibase https://doi.org/10.1016/j.cpc.2017.09.033} {\bibfield  {journal} {\bibinfo  {journal} {Computer Physics Communications}\ }\textbf {\bibinfo {volume} {224}},\ \bibinfo {pages} {405} (\bibinfo {year} {2018})}\BibitemShut {NoStop}%
\bibitem [{\citenamefont {Zhang}\ \emph {et~al.}(2022)\citenamefont {Zhang}, \citenamefont {Yu}, \citenamefont {Liu},\ and\ \citenamefont {Yao}}]{Zhang2022}%
  \BibitemOpen
  \bibfield  {author} {\bibinfo {author} {\bibfnamefont {Z.}~\bibnamefont {Zhang}}, \bibinfo {author} {\bibfnamefont {Z.}~\bibnamefont {Yu}}, \bibinfo {author} {\bibfnamefont {G.}~\bibnamefont {Liu}}, \ and\ \bibinfo {author} {\bibfnamefont {Y.}~\bibnamefont {Yao}},\ }\href {\doibase https://doi.org/10.1016/j.cpc.2021.108153} {\bibfield  {journal} {\bibinfo  {journal} {Computer Physics Communications}\ }\textbf {\bibinfo {volume} {270}},\ \bibinfo {pages} {108153} (\bibinfo {year} {2022})}\BibitemShut {NoStop}%
\bibitem [{\citenamefont {Korm\'anyos}\ \emph {et~al.}(2013)\citenamefont {Korm\'anyos}, \citenamefont {Z\'olyomi}, \citenamefont {Drummond}, \citenamefont {Rakyta}, \citenamefont {Burkard},\ and\ \citenamefont {Fal'ko}}]{Falko2013}%
  \BibitemOpen
  \bibfield  {author} {\bibinfo {author} {\bibfnamefont {A.}~\bibnamefont {Korm\'anyos}}, \bibinfo {author} {\bibfnamefont {V.}~\bibnamefont {Z\'olyomi}}, \bibinfo {author} {\bibfnamefont {N.~D.}\ \bibnamefont {Drummond}}, \bibinfo {author} {\bibfnamefont {P.}~\bibnamefont {Rakyta}}, \bibinfo {author} {\bibfnamefont {G.}~\bibnamefont {Burkard}}, \ and\ \bibinfo {author} {\bibfnamefont {V.~I.}\ \bibnamefont {Fal'ko}},\ }\href {\doibase 10.1103/PhysRevB.88.045416} {\bibfield  {journal} {\bibinfo  {journal} {Phys. Rev. B}\ }\textbf {\bibinfo {volume} {88}},\ \bibinfo {pages} {045416} (\bibinfo {year} {2013})}\BibitemShut {NoStop}%
\bibitem [{\citenamefont {Liu}\ \emph {et~al.}(2013)\citenamefont {Liu}, \citenamefont {Shan}, \citenamefont {Yao}, \citenamefont {Yao},\ and\ \citenamefont {Xiao}}]{Liu2013}%
  \BibitemOpen
  \bibfield  {author} {\bibinfo {author} {\bibfnamefont {G.-B.}\ \bibnamefont {Liu}}, \bibinfo {author} {\bibfnamefont {W.-Y.}\ \bibnamefont {Shan}}, \bibinfo {author} {\bibfnamefont {Y.}~\bibnamefont {Yao}}, \bibinfo {author} {\bibfnamefont {W.}~\bibnamefont {Yao}}, \ and\ \bibinfo {author} {\bibfnamefont {D.}~\bibnamefont {Xiao}},\ }\href {\doibase 10.1103/PhysRevB.88.085433} {\bibfield  {journal} {\bibinfo  {journal} {Phys. Rev. B}\ }\textbf {\bibinfo {volume} {88}},\ \bibinfo {pages} {085433} (\bibinfo {year} {2013})}\BibitemShut {NoStop}%
\bibitem [{\citenamefont {Ribeiro}\ \emph {et~al.}(2009)\citenamefont {Ribeiro}, \citenamefont {Pereira}, \citenamefont {Peres}, \citenamefont {Briddon},\ and\ \citenamefont {Neto}}]{Ribeiro2009}%
  \BibitemOpen
  \bibfield  {author} {\bibinfo {author} {\bibfnamefont {R.~M.}\ \bibnamefont {Ribeiro}}, \bibinfo {author} {\bibfnamefont {V.~M.}\ \bibnamefont {Pereira}}, \bibinfo {author} {\bibfnamefont {N.~M.~R.}\ \bibnamefont {Peres}}, \bibinfo {author} {\bibfnamefont {P.~R.}\ \bibnamefont {Briddon}}, \ and\ \bibinfo {author} {\bibfnamefont {A.~H.~C.}\ \bibnamefont {Neto}},\ }\href {\doibase 10.1088/1367-2630/11/11/115002} {\bibfield  {journal} {\bibinfo  {journal} {New Journal of Physics}\ }\textbf {\bibinfo {volume} {11}},\ \bibinfo {pages} {115002} (\bibinfo {year} {2009})}\BibitemShut {NoStop}%
\bibitem [{\citenamefont {Qiao}\ \emph {et~al.}(2010)\citenamefont {Qiao}, \citenamefont {Yang}, \citenamefont {Feng}, \citenamefont {Tse}, \citenamefont {Ding}, \citenamefont {Yao}, \citenamefont {Wang},\ and\ \citenamefont {Niu}}]{Qiao2010}%
  \BibitemOpen
  \bibfield  {author} {\bibinfo {author} {\bibfnamefont {Z.}~\bibnamefont {Qiao}}, \bibinfo {author} {\bibfnamefont {S.~A.}\ \bibnamefont {Yang}}, \bibinfo {author} {\bibfnamefont {W.}~\bibnamefont {Feng}}, \bibinfo {author} {\bibfnamefont {W.-K.}\ \bibnamefont {Tse}}, \bibinfo {author} {\bibfnamefont {J.}~\bibnamefont {Ding}}, \bibinfo {author} {\bibfnamefont {Y.}~\bibnamefont {Yao}}, \bibinfo {author} {\bibfnamefont {J.}~\bibnamefont {Wang}}, \ and\ \bibinfo {author} {\bibfnamefont {Q.}~\bibnamefont {Niu}},\ }\href {\doibase 10.1103/PhysRevB.82.161414} {\bibfield  {journal} {\bibinfo  {journal} {Phys. Rev. B}\ }\textbf {\bibinfo {volume} {82}},\ \bibinfo {pages} {161414} (\bibinfo {year} {2010})}\BibitemShut {NoStop}%
\bibitem [{\citenamefont {Xu}\ \emph {et~al.}(2011)\citenamefont {Xu}, \citenamefont {Weng}, \citenamefont {Wang}, \citenamefont {Dai},\ and\ \citenamefont {Fang}}]{Xu2011}%
  \BibitemOpen
  \bibfield  {author} {\bibinfo {author} {\bibfnamefont {G.}~\bibnamefont {Xu}}, \bibinfo {author} {\bibfnamefont {H.}~\bibnamefont {Weng}}, \bibinfo {author} {\bibfnamefont {Z.}~\bibnamefont {Wang}}, \bibinfo {author} {\bibfnamefont {X.}~\bibnamefont {Dai}}, \ and\ \bibinfo {author} {\bibfnamefont {Z.}~\bibnamefont {Fang}},\ }\href {\doibase 10.1103/PhysRevLett.107.186806} {\bibfield  {journal} {\bibinfo  {journal} {Phys. Rev. Lett.}\ }\textbf {\bibinfo {volume} {107}},\ \bibinfo {pages} {186806} (\bibinfo {year} {2011})}\BibitemShut {NoStop}%
\bibitem [{\citenamefont {Fang}\ \emph {et~al.}(2012)\citenamefont {Fang}, \citenamefont {Gilbert}, \citenamefont {Dai},\ and\ \citenamefont {Bernevig}}]{Fang2012}%
  \BibitemOpen
  \bibfield  {author} {\bibinfo {author} {\bibfnamefont {C.}~\bibnamefont {Fang}}, \bibinfo {author} {\bibfnamefont {M.~J.}\ \bibnamefont {Gilbert}}, \bibinfo {author} {\bibfnamefont {X.}~\bibnamefont {Dai}}, \ and\ \bibinfo {author} {\bibfnamefont {B.~A.}\ \bibnamefont {Bernevig}},\ }\href {\doibase 10.1103/PhysRevLett.108.266802} {\bibfield  {journal} {\bibinfo  {journal} {Phys. Rev. Lett.}\ }\textbf {\bibinfo {volume} {108}},\ \bibinfo {pages} {266802} (\bibinfo {year} {2012})}\BibitemShut {NoStop}%
\bibitem [{\citenamefont {Wang}\ \emph {et~al.}(2013)\citenamefont {Wang}, \citenamefont {Liu},\ and\ \citenamefont {Liu}}]{Wang2013}%
  \BibitemOpen
  \bibfield  {author} {\bibinfo {author} {\bibfnamefont {Z.~F.}\ \bibnamefont {Wang}}, \bibinfo {author} {\bibfnamefont {Z.}~\bibnamefont {Liu}}, \ and\ \bibinfo {author} {\bibfnamefont {F.}~\bibnamefont {Liu}},\ }\href {\doibase 10.1103/PhysRevLett.110.196801} {\bibfield  {journal} {\bibinfo  {journal} {Phys. Rev. Lett.}\ }\textbf {\bibinfo {volume} {110}},\ \bibinfo {pages} {196801} (\bibinfo {year} {2013})}\BibitemShut {NoStop}%
\bibitem [{\citenamefont {He}\ \emph {et~al.}(2017)\citenamefont {He}, \citenamefont {Li}, \citenamefont {Lyu},\ and\ \citenamefont {Nachtigall}}]{He2017}%
  \BibitemOpen
  \bibfield  {author} {\bibinfo {author} {\bibfnamefont {J.}~\bibnamefont {He}}, \bibinfo {author} {\bibfnamefont {X.}~\bibnamefont {Li}}, \bibinfo {author} {\bibfnamefont {P.}~\bibnamefont {Lyu}}, \ and\ \bibinfo {author} {\bibfnamefont {P.}~\bibnamefont {Nachtigall}},\ }\href {\doibase 10.1039/C6NR08522A} {\bibfield  {journal} {\bibinfo  {journal} {Nanoscale}\ }\textbf {\bibinfo {volume} {9}},\ \bibinfo {pages} {2246} (\bibinfo {year} {2017})}\BibitemShut {NoStop}%
\bibitem [{\citenamefont {Li}\ \emph {et~al.}(2017)\citenamefont {Li}, \citenamefont {Li}, \citenamefont {Zhao}, \citenamefont {Chen}, \citenamefont {Chen}, \citenamefont {Guo}, \citenamefont {Feng}, \citenamefont {Gong},\ and\ \citenamefont {MacDonald}}]{Li2017}%
  \BibitemOpen
  \bibfield  {author} {\bibinfo {author} {\bibfnamefont {P.}~\bibnamefont {Li}}, \bibinfo {author} {\bibfnamefont {X.}~\bibnamefont {Li}}, \bibinfo {author} {\bibfnamefont {W.}~\bibnamefont {Zhao}}, \bibinfo {author} {\bibfnamefont {H.}~\bibnamefont {Chen}}, \bibinfo {author} {\bibfnamefont {M.-X.}\ \bibnamefont {Chen}}, \bibinfo {author} {\bibfnamefont {Z.-X.}\ \bibnamefont {Guo}}, \bibinfo {author} {\bibfnamefont {J.}~\bibnamefont {Feng}}, \bibinfo {author} {\bibfnamefont {X.-G.}\ \bibnamefont {Gong}}, \ and\ \bibinfo {author} {\bibfnamefont {A.~H.}\ \bibnamefont {MacDonald}},\ }\href {\doibase 10.1021/acs.nanolett.7b02855} {\bibfield  {journal} {\bibinfo  {journal} {Nano Lett.}\ }\textbf {\bibinfo {volume} {17}},\ \bibinfo {pages} {6195} (\bibinfo {year} {2017})},\ \bibinfo {note} {publisher: American Chemical Society}\BibitemShut {NoStop}%
\bibitem [{\citenamefont {Li}\ \emph {et~al.}(2019)\citenamefont {Li}, \citenamefont {Zhang}, \citenamefont {Zhao}, \citenamefont {Xue},\ and\ \citenamefont {Yang}}]{Yang2019}%
  \BibitemOpen
  \bibfield  {author} {\bibinfo {author} {\bibfnamefont {Y.}~\bibnamefont {Li}}, \bibinfo {author} {\bibfnamefont {J.}~\bibnamefont {Zhang}}, \bibinfo {author} {\bibfnamefont {B.}~\bibnamefont {Zhao}}, \bibinfo {author} {\bibfnamefont {Y.}~\bibnamefont {Xue}}, \ and\ \bibinfo {author} {\bibfnamefont {Z.}~\bibnamefont {Yang}},\ }\href {\doibase 10.1103/PhysRevB.99.195402} {\bibfield  {journal} {\bibinfo  {journal} {Phys. Rev. B}\ }\textbf {\bibinfo {volume} {99}},\ \bibinfo {pages} {195402} (\bibinfo {year} {2019})}\BibitemShut {NoStop}%
\end{thebibliography}
\end{document}